# *Synthetic Axion Response with Spacetime Crystals*

**Filipa R. Prudêncio[1,2*], and Mário G. Silveirinha[1]**

[1]University of Lisbon – Instituto Superior Técnico and Instituto de Telecomunicações,

Avenida Rovisco Pais 1, 1049-001 Lisbon, Portugal

[2]Instituto Universitário de Lisboa (ISCTE-IUL), Avenida das Forças Armadas 376, 1600-

077 Lisbon, Portugal

## Abstract

Here, we show that spacetime modulations provide an exciting route to realize complex nonreciprocal couplings, and in particular the elusive axion response. We develop an analytical formalism to homogenize anisotropic spacetime crystals in the long wavelength limit. It is found that spacetime crystals with suitable glide-rotation symmetry can have a giant axion-type response, several orders of magnitude larger than in natural materials. The nonreciprocal axion response may have interesting applications in optics, for example in electromagnetic isolation, and in addition may enable exciting new forms of light-wave interactions.

[*] Corresponding author: filipa.prudencio@lx.it.pt

# I. Introduction

With the rise of metamaterials, the "toolbox" of available electromagnetic responses was greatly expanded in the past two decades. Today, the principles that guide the design of media with rather exotic properties, e.g., a double negative permittivity and permeability [1], extreme anisotropy [2], hyperbolic isofrequency contours [3, 4], topological systems immune to back-scattering relying on a duality symmetry [5-7], and others, are well understood. There is however a rather peculiar class of materials for which limited progress has been achieved. That is the class of media with an axion or Tellegen coupling [8-12]. In its simpler isotropic version, such materials are characterized by a constitutive relation of the form:

$$\mathbf{D} = \varepsilon_0 \varepsilon \mathbf{E} + \frac{\kappa}{c} \mathbf{H}, \qquad \mathbf{B} = \frac{\kappa}{c} \mathbf{E} + \mu_0 \mu \mathbf{H}, \qquad (1)$$

where $\kappa$ is the Tellegen parameter. Here, $\mathbf{E}$ is the electric field, $\mathbf{H}$ is the magnetic field, $\mathbf{D}$ is the electric displacement and $\mathbf{B}$ is the magnetic induction field. The axion medium is nonreciprocal, but different from typical nonreciprocal gyrotropic materials (e.g., ferrites and other iron garnets), the nonreciprocity does not necessarily require an external bias and may be anchored in a spontaneous time-symmetry breaking [11], e.g., due to anti-ferromagnetism [12-14].

Remarkably the electromagnetic response of an isotropic Tellegen medium can be linked through a duality transformation to the response of conventional dielectrics [9, 15]. Therefore, the dispersion of waves in Tellegen materials is in some cases indistinguishable from the dispersion in conventional structures [9, 10, 12]. This property has led to some scientific debate on the physical viability of the axion response [16-22]. Today, this debate is settled, and it is widely accepted that axion-type responses can emerge spontaneously –albeit extremely weak – in nature in some antiferromagnetic materials [13], and most notably in electronic topological insulators [14, 23, 24, 25]. Furthermore, the elusive axion response is interesting also from a conceptual point of view due its connections with high-energy physics. It was suggested by Wilczek that "axions" may provide an explanation for dark matter [26].

As mentioned previously, the Tellegen parameter $\kappa$ of naturally available materials is usually negligibly small. For example, for $Cr_2O_3$ it is on the order of $10^{-4}$ [13]. Due to this reason the signature of Tellegen induced electromagnetic phenomena is rather faint, and can be ignored for all practical purposes. Tellegen materials can be in principle synthesized artificially. A proposal relying on a gyrotropic magnetic particle is reported in Ref. [27]. A related idea relying on an electrically-biased gyrotropic particle [28] is described in Ref. [29].

Recently, the time modulation of the material parameters has opened up new opportunities in the design of different classes of metamaterials [30-39]. For example, since the time modulation breaks the Lorentz reciprocity in electromagnetism, it has been shown that time-varying material responses can be used to realize unidirectional guides and isolators, multifunctional non-reciprocal metasurfaces, and others, without an external magnetic bias [30-37]. Moreover, rather remarkably, time modulated photonic crystals can be used as a platform to synthesize exotic metamaterial responses [40-43]. In particular, it was shown in Ref. [43] that spacetime crystals with a travelling wave-type spacetime modulation behave in many ways as a synthetic moving medium. In such systems the permittivity and permeability depend on the spacetime coordinates as $\varepsilon = \varepsilon(z - vt)$ and $\mu = \mu(z - vt)$, with $v$ the modulation speed. This type of modulation leads to a synthetic Fresnel drag effect such that the electromagnetic waves are dragged by the synthetic motion of the medium, in the same manner as they would be dragged by a moving material body [41]. Moreover, in the long wavelength limit, the spacetime crystal can be described using effective parameters, which reproduce the well-known moving medium coupling [41, 43, 44]. The velocity of the equivalent moving medium is different from the modulation velocity *v*, and, curiously, its sign may also differ from the sign of *v*.

In view of these developments, it is natural to wonder if it may be possible to implement the axion-coupling using a spacetime modulated crystal. Here, using homogenization theory and full-wave simulations, we demonstrate that, indeed, spacetime modulated systems offer a unique opportunity to realize a metamaterial with a giant axion-type response.

The article is organized as follows. In section II, we extend the homogenization theory of Ref. [43] to anisotropic spacetime crystals with a travelling wave-type spacetime modulation. Based on symmetry considerations, we propose a general solution to design an effective medium that behaves as a moving-Tellegen medium for propagation along the direction of the modulation velocity. In particular, we present a simple solution to implement the synthetic axion response. In Section III, we present a detailed study of the effective response of the synthetic axion material. In addition, we illustrate the impact of the synthetic axion coupling by studying the reflection of a plane wave at an interface between air and the effective material. It is shown that the spacetime modulation can result in a giant rotation of polarization of the incident wave. Finally, section IV summarizes the main conclusions.

## II. HOMOGENIZATION OF ANISOTROPIC SPACETIME SYSTEMS

We consider a generic one-dimensional spacetime crystal with constitutive relations of the type

$$\mathbf{D}(x,y,z,t) = \varepsilon_0 \bar{\varepsilon}(z-vt) \cdot \mathbf{E}(x,y,z,t), \quad \mathbf{B}(x,y,z,t) = \mu_0 \bar{\mu}(z-vt) \cdot \mathbf{H}(x,y,z,t). \quad (2)$$

The material parameters vary both in space and in time, but are constant for spacetime points such that $z' = z - vt = const.$, being $v$ the modulation speed. This is known as a travelling wave modulation, and different aspects of it have been studied by several authors in recent years [34, 35, 41, 42, 43] (for earlier works see [45, 46]). Here, we suppose that the permittivity and the permeability are anisotropic and described by real-valued symmetric periodic tensors of the type:

$$\bar{\varepsilon} = \begin{pmatrix} \varepsilon_{xx} & \varepsilon_{xy} & 0 \\ \varepsilon_{xy} & \varepsilon_{yy} & 0 \\ 0 & 0 & \varepsilon_{zz} \end{pmatrix}, \qquad \bar{\mu} = \begin{pmatrix} \mu_{xx} & \mu_{xy} & 0 \\ \mu_{xy} & \mu_{yy} & 0 \\ 0 & 0 & \mu_{zz} \end{pmatrix}, \qquad (3)$$

with $\bar{\varepsilon}(z') = \bar{\varepsilon}(z'+a)$ and $\bar{\mu}(z') = \bar{\mu}(z'+a)$ periodic functions with a spatial period $a$ (see Fig. 1). Time dependent anisotropic materials can enable rather peculiar electromagnetic phenomena [47-48].

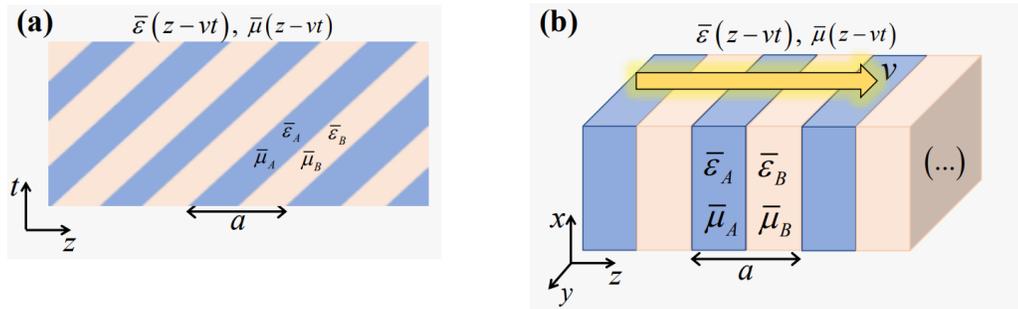

**Fig. 1** Anisotropic spacetime crystal formed by two layers with material tensors $\bar{\varepsilon}_A, \bar{\mu}_A$ and $\bar{\varepsilon}_B, \bar{\mu}_B$. **(a)** Representation of the material parameters in a spacetime diagram. **(b)** Sketch of the stratified crystal in real space with the arrow indicating how the material parameters vary in time. The modulation speed is $\mathbf{v} = v\hat{\mathbf{z}}$.

Due to the assumed structure of $\bar{\varepsilon}$ and $\bar{\mu}$ [Eq. (3)], the directions that diagonalize each of the tensors (loosely referred to "optical axes") are such that: (i) two optical axes are in the *xoy* plane and (ii) the third optical axis is along *z*. Recently, it has been shown that by changing continuously the optical axes it is possible to realize an Archimedes screw for light [42]. In

–5–

contrast, in the present work the optical axes are assumed fixed, i.e., independent of space and time. However, the permittivity and permeability tensors in general may not share the same optical axes in the *xoy* plane (see Fig. 2a). The angular offset between the permeability and permittivity optical axes is denoted by $\theta$. It will be shown in Sect. III that a nontrivial $\theta$ is the key to unlock the axion response.

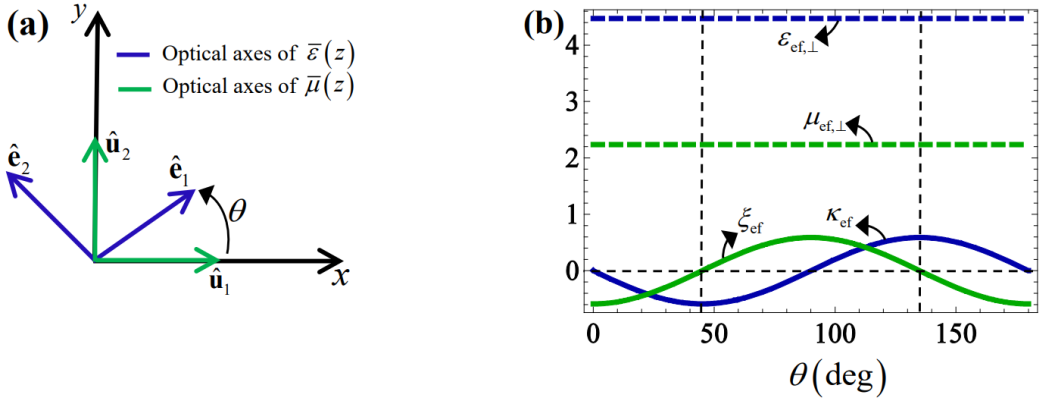

**Fig. 2 (a)** Optical axes of the permittivity and permeability tensors, $\bar{\varepsilon}(z)$ and $\bar{\mu}(z)$, respectively. The offset angle is denoted by $\theta$. **(b)** Parameters of the effective moving-Tellegen medium, $\varepsilon_{\text{ef},\perp}, \mu_{\text{ef},\perp}, \kappa_{\text{ef}}, \xi_{\text{ef}}$, as a function of the angle $\theta$, for $\varepsilon_1 = 6$, $\varepsilon_2 = 2$, $\mu_1 = 3$ and $\mu_2 = 1$. An effective moving medium is obtained for $\theta = 0, 90°, 180°$, whereas for $\theta = 45°, 135°$ (dashed vertical grid lines) a pure axion response is achieved. The modulation speed is $v = 0.2c$.

For future reference, we introduce a 4×4 matrix that describes the transverse material response of the spacetime crystal:

$$\mathbf{M}_\perp(z') = \begin{pmatrix} \varepsilon_0 \bar{\varepsilon}_\perp & \mathbf{0}_{2\times 2} \\ \mathbf{0}_{2\times 2} & \mu_0 \bar{\mu}_\perp \end{pmatrix}, \qquad \text{with} \tag{4a}$$

$$\bar{\varepsilon}_\perp = \begin{pmatrix} \varepsilon_{xx}(z') & \varepsilon_{xy}(z') \\ \varepsilon_{xy}(z') & \varepsilon_{yy}(z') \end{pmatrix}, \quad \text{and} \quad \bar{\mu}_\perp = \begin{pmatrix} \mu_{xx}(z') & \mu_{xy}(z') \\ \mu_{xy}(z') & \mu_{yy}(z') \end{pmatrix}. \tag{4b}$$

## A. *General homogenization theory*

The homogenization of the spacetime crystal is done using the same strategy as in Ref. [43]. First, we use a Galilean coordinate transformation ($\mathbf{r}' = \mathbf{r} - \mathbf{v}t$, $t' = t$) to switch to a frame (the "co-moving frame") where the material parameters become independent of time. Then, the crystal response is homogenized in the co-moving frame. Finally, we use the inverse Galilean transformation to determine the effective response in the original "laboratory" frame.

To begin with, we write the Maxwell's equations in the lab frame in the following compact manner:

$$\underbrace{\begin{pmatrix} \mathbf{0} & i\nabla \times \mathbf{1}_{3\times 3} \\ -i\nabla \times \mathbf{1}_{3\times 3} & \mathbf{0} \end{pmatrix}}_{\hat{N}(-i\nabla)} \cdot \begin{pmatrix} \mathbf{E} \\ \mathbf{H} \end{pmatrix} = i\frac{\partial}{\partial t}\begin{pmatrix} \mathbf{D} \\ \mathbf{B} \end{pmatrix}, \tag{5a}$$

$$\begin{pmatrix} \mathbf{D} \\ \mathbf{B} \end{pmatrix} = \begin{pmatrix} \overline{\varepsilon}(z-v_0 t) & \mathbf{0} \\ \mathbf{0} & \overline{\mu}(z-v_0 t) \end{pmatrix} \cdot \begin{pmatrix} \mathbf{E} \\ \mathbf{H} \end{pmatrix} \equiv \mathbf{M}(z-v_0 t) \cdot \begin{pmatrix} \mathbf{E} \\ \mathbf{H} \end{pmatrix}. \tag{5b}$$

The structure of the above equations is preserved under a Galilean transformation (or more generally under a Lorentz transformation). Therefore, the fields in the co-moving frame satisfy

$$\hat{N}(-i\nabla') \cdot \begin{pmatrix} \mathbf{E}' \\ \mathbf{H}' \end{pmatrix} = i\frac{\partial}{\partial t}\begin{pmatrix} \mathbf{D}' \\ \mathbf{B}' \end{pmatrix} \quad \text{with} \quad \begin{pmatrix} \mathbf{D}' \\ \mathbf{B}' \end{pmatrix} = \mathbf{M}'(z') \cdot \begin{pmatrix} \mathbf{E}' \\ \mathbf{H}' \end{pmatrix}.$$

Here, $\mathbf{M}'(z')$ is the (time-independent) material matrix in the co-moving frame, which is related to the unprimed material matrix as (see Appendix A, Eq. (A8)):

$$\mathbf{M}' = \mathbf{M} \cdot [\mathbf{1}_{6\times 6} + \mathbf{V} \cdot \mathbf{M}]^{-1}, \quad \text{with} \quad \mathbf{V} = \begin{pmatrix} \mathbf{0}_{3\times 3} & \mathbf{v} \times \mathbf{1}_{3\times 3} \\ -\mathbf{v} \times \mathbf{1}_{3\times 3} & \mathbf{0}_{3\times 3} \end{pmatrix}. \tag{6}$$

In the above, $\mathbf{v} = v\hat{\mathbf{z}}$ is the modulation speed.

It is shown in Appendix B that the standard homogenization approach for stratified structures can be readily generalized to anisotropic spacetime crystals. The effective material matrix in the

–7–

lab frame $\mathbf{M}_{\text{ef}}$ links the spatially averaged fields as follows $\begin{pmatrix} \langle \mathbf{D} \rangle \\ \langle \mathbf{B} \rangle \end{pmatrix} = \mathbf{M}_{\text{ef}} \cdot \begin{pmatrix} \langle \mathbf{E} \rangle \\ \langle \mathbf{H} \rangle \end{pmatrix}$. For "microscopic" electric and magnetic responses as in Eq. (3), the effective material response is such that:

$$\mathbf{M}_{\text{ef}} = \begin{pmatrix} \varepsilon_0 \overline{\overline{\varepsilon}}_{\text{ef}} & \frac{1}{c} \overline{\overline{\xi}}_{\text{ef}} \\ \frac{1}{c} \overline{\overline{\zeta}}_{\text{ef}} & \mu_0 \overline{\overline{\mu}}_{\text{ef}} \end{pmatrix}, \quad \text{with} \tag{7a}$$

$$\overline{\overline{\varepsilon}}_{\text{ef}} = \begin{pmatrix} \varepsilon_{\text{ef},xx} & \varepsilon_{\text{ef},xy} & 0 \\ \varepsilon_{\text{ef},xy} & \varepsilon_{\text{ef},yy} & 0 \\ 0 & 0 & \varepsilon_{\text{ef},zz} \end{pmatrix}, \quad \overline{\overline{\mu}}_{\text{ef}} = \begin{pmatrix} \mu_{\text{ef},xx} & \mu_{\text{ef},xy} & 0 \\ \mu_{\text{ef},xy} & \mu_{\text{ef},yy} & 0 \\ 0 & 0 & \mu_{\text{ef},zz} \end{pmatrix}, \tag{7b}$$

$$\overline{\overline{\xi}}_{\text{ef}} = \overline{\overline{\zeta}}_{\text{ef}}^T = \begin{pmatrix} \xi_{\text{ef},xx} & \xi_{\text{ef},xy} & 0 \\ \xi_{\text{ef},yx} & \xi_{\text{ef},yy} & 0 \\ 0 & 0 & 0 \end{pmatrix}. \tag{7c}$$

The general formula for the effective material parameters is given in Appendix B [Eqs. (B4)-(B5)]. The effective parameters $\varepsilon_{\text{ef},zz}$ and $\mu_{\text{ef},zz}$ can be expressed in terms of $\varepsilon_{zz}(z')$ and $\mu_{zz}(z')$ as:

$$\varepsilon_{\text{ef},zz} = \left[ \frac{1}{a} \int_0^a \frac{1}{\varepsilon_{zz}(z')} dz' \right]^{-1} \text{ and } \mu_{\text{ef},zz} = \left[ \frac{1}{a} \int_0^a \frac{1}{\mu_{zz}(z')} dz' \right]^{-1}. \tag{8}$$

Note that $\varepsilon_{\text{ef},zz}$ and $\mu_{\text{ef},zz}$ are independent of the modulation speed. Indeed, in the long wavelength limit, the response of the crystal along the $z$ direction (parallel to the modulation speed) is effectively decoupled from the response of the material in the *xoy* plane (see Appendix A).

Since the response in the *xoy* plane is decoupled from the response along the *z*-direction, it is possible to write the remaining effective parameters in terms of $\mathbf{M}_\perp$ [Eq. (4)]. To this end, it is convenient to introduce the 4×4 matrix

$$\mathbf{M}_{\text{ef},\perp} = \begin{pmatrix} \varepsilon_0 \bar{\varepsilon}_{\text{ef},\perp} & \dfrac{1}{c} \bar{\xi}_{\text{ef},\perp} \\ \dfrac{1}{c} \bar{\zeta}_{\text{ef},\perp} & \mu_0 \bar{\mu}_{\text{ef},\perp} \end{pmatrix} \tag{9}$$

where $\bar{\varepsilon}_{\text{ef},\perp} \equiv \begin{pmatrix} \varepsilon_{\text{ef},xx} & \varepsilon_{\text{ef},xy} \\ \varepsilon_{\text{ef},xy} & \varepsilon_{\text{ef},yy} \end{pmatrix}$, $\bar{\mu}_{\text{ef},\perp} \equiv \begin{pmatrix} \mu_{\text{ef},xx} & \mu_{\text{ef},xy} \\ \mu_{\text{ef},xy} & \mu_{\text{ef},yy} \end{pmatrix}$, and $\bar{\xi}_{\text{ef},\perp} = \left( \bar{\zeta}_{\text{ef},\perp} \right)^T \equiv \begin{pmatrix} \xi_{\text{ef},xx} & \xi_{\text{ef},xy} \\ \xi_{\text{ef},yx} & \xi_{\text{ef},yy} \end{pmatrix}$ are

2×2 matrices that determine the effective response of the spacetime crystal in the *xoy* plane [Eq. (7)]. Using Eqs. (B4)-(B5) one may show that:

$$\mathbf{M}_{\text{ef},\perp} = \mathbf{M}'_{\text{ef},\perp} \cdot \left[ \mathbf{1}_{4 \times 4} - v \boldsymbol{\sigma} \cdot \mathbf{M}'_{\text{ef},\perp} \right]^{-1}, \quad \text{with} \tag{10a}$$

$$\mathbf{M}'_{\text{ef},\perp} = \frac{1}{a} \int_0^a dz' \, \mathbf{M}_\perp(z') \cdot \left[ \mathbf{1}_{4 \times 4} + v \boldsymbol{\sigma} \cdot \mathbf{M}_\perp(z') \right]^{-1}, \tag{10b}$$

$$\boldsymbol{\sigma} = \begin{pmatrix} \mathbf{0}_{2 \times 2} & \mathbf{J} \\ -\mathbf{J} & \mathbf{0}_{2 \times 2} \end{pmatrix}, \quad \text{with} \quad \mathbf{J} = \begin{pmatrix} 0 & -1 \\ 1 & 0 \end{pmatrix}. \tag{10c}$$

Note that $\mathbf{M}_{\text{ef},\perp}, \mathbf{M}'_{\text{ef},\perp}, \mathbf{M}_\perp, \boldsymbol{\sigma}$ are all 4×4 real-valued matrices. In general, the effective response in the lab frame is bianisotropic so that the magneto-electric tensors $\bar{\xi}_{\text{ef}}, \bar{\zeta}_{\text{ef}}$ are nontrivial. Analogous to Ref. [43], it can be shown that a nontrivial bianisotropic response can occur only when both the permittivity and permeability tensors are simultaneously modulated in space and in time.

## B. Glide-rotation symmetry

Next, we show that by enforcing the invariance of the spacetime crystal under a particular glide symmetry, it is possible reduce drastically the complexity of the effective response. The glide symmetry ensures that the material response is invariant under continuous rotations about the $z$-axis, i.e. independent of the field polarization in the $xoy$ plane.

It is convenient to write the transverse permittivity and permeability tensors [Eq. (4)] as a function of its eigenvalues. As shown in Fig. 2a, it is supposed without loss of generality that the principal axes of the permeability are aligned with the $x$ and $y$ directions: $\hat{\mathbf{u}}_1 \equiv \hat{\mathbf{x}}, \hat{\mathbf{u}}_2 \equiv \hat{\mathbf{y}}$. The principal axes of the permittivity tensor $\hat{\mathbf{e}}_1, \hat{\mathbf{e}}_2$ are tilted by $\theta$ with respect to the axes of the permeability. These properties imply that: $\overline{\mu}_\perp(z,t) = \mu_1(z')\hat{\mathbf{u}}_1 \otimes \hat{\mathbf{u}}_1 + \mu_2(z')\hat{\mathbf{u}}_2 \otimes \hat{\mathbf{u}}_2$ and $\overline{\varepsilon}_\perp(z,t) = \varepsilon_1(z')\hat{\mathbf{e}}_1 \otimes \hat{\mathbf{e}}_1 + \varepsilon_2(z')\hat{\mathbf{e}}_2 \otimes \hat{\mathbf{e}}_2$ with $z' = z - vt$ and $\mu_1, \mu_2$ ($\varepsilon_1, \varepsilon_2$) the eigenvalues of the permeability (permittivity). The two tensors can be explicitly written in the $xoy$ coordinate system as:

$$\overline{\varepsilon}_\perp(z,t) = \begin{pmatrix} \varepsilon_1(z')\cos^2\theta + \varepsilon_2(z')\sin^2\theta & [\varepsilon_1(z') - \varepsilon_2(z')]\cos\theta\sin\theta \\ [\varepsilon_1(z') - \varepsilon_2(z')]\cos\theta\sin\theta & \varepsilon_1(z')\sin^2\theta + \varepsilon_2(z')\cos^2\theta \end{pmatrix}, \quad (11a)$$

$$\overline{\mu}_\perp(z,t) = \begin{pmatrix} \mu_1(z') & 0 \\ 0 & \mu_2(z') \end{pmatrix}. \quad (11b)$$

Let us now suppose that eigenvalues of each material tensor are linked as follows:

$$\varepsilon_2(z') = \varepsilon_1\left(z' - \frac{a}{2}\right), \qquad \mu_2(z') = \mu_1\left(z' - \frac{a}{2}\right). \quad (12)$$

In such a case, the microscopic (transverse) material matrix $\mathbf{M}_\perp$ [Eq. (4a)] is invariant under a rotation of 90º about the z-axis followed by a translation by half lattice constant:

–10–

$$\mathbf{M}_{\perp}(z') = \begin{pmatrix} \mathbf{J}^T & \mathbf{0}_{2\times 2} \\ \mathbf{0}_{2\times 2} & \mathbf{J}^T \end{pmatrix} \cdot \mathbf{M}_{\perp}\left(z' - \frac{a}{2}\right) \cdot \begin{pmatrix} \mathbf{J} & \mathbf{0}_{2\times 2} \\ \mathbf{0}_{2\times 2} & \mathbf{J} \end{pmatrix}, \tag{13}$$

with $\mathbf{J} = \begin{pmatrix} 0 & -1 \\ 1 & 0 \end{pmatrix}$ the 90º-rotation matrix introduced in subsection III.A. We will refer to this transformation symmetry as a *glide-rotation symmetry* (see Ref. [49] for an overview of glide symmetries). Evidently, the effective medium theory must preserve the glide-rotation symmetry. As the homogenized response is independent of the spacetime coordinates, it follows that:

$$\mathbf{M}_{\perp,\text{ef}} = \begin{pmatrix} \mathbf{J}^T & \mathbf{0}_{2\times 2} \\ \mathbf{0}_{2\times 2} & \mathbf{J}^T \end{pmatrix} \cdot \mathbf{M}_{\perp,\text{ef}} \cdot \begin{pmatrix} \mathbf{J} & \mathbf{0}_{2\times 2} \\ \mathbf{0}_{2\times 2} & \mathbf{J} \end{pmatrix}. \tag{14}$$

Substituting now Eq. (9) into the above formula, it is found that the enforced symmetry implies that: $\bar{\varepsilon}_{\text{ef},\perp} = \mathbf{J}^T \cdot \bar{\varepsilon}_{\text{ef},\perp} \cdot \mathbf{J}$, $\bar{\mu}_{\text{ef},\perp} = \mathbf{J}^T \cdot \bar{\mu}_{\text{ef},\perp} \cdot \mathbf{J}$, and $\bar{\xi}_{\text{ef},\perp} = \mathbf{J}^T \cdot \bar{\xi}_{\text{ef},\perp} \cdot \mathbf{J}$. Since $\bar{\varepsilon}_{\text{ef},\perp}, \bar{\mu}_{\text{ef},\perp}$ are symmetric tensors they must be scalars: $\bar{\varepsilon}_{\text{ef},\perp} = \varepsilon_{\text{ef},\perp} \mathbf{1}_{\perp}$ and $\bar{\mu}_{\text{ef},\perp} = \mu_{\text{ef},\perp} \mathbf{1}_{\perp}$. On the other hand, the magneto-electric tensors are determined by two scalar parameters $\kappa_{\text{ef}}, \xi_{\text{ef}}$ such that $\bar{\xi}_{\text{ef},\perp} = \kappa_{\text{ef}} \mathbf{1}_{\perp} - \xi_{\text{ef}} \mathbf{J}$ and $\bar{\zeta}_{\text{ef},\perp} = \left(\bar{\xi}_{\text{ef},\perp}\right)^T = \kappa_{\text{ef}} \mathbf{1}_{\perp} + \xi_{\text{ef}} \mathbf{J}$. Note that $\kappa_{\text{ef}} = \xi_{xx,\text{ef}} = \xi_{yy,\text{ef}}$ and $\xi_{\text{ef}} = \xi_{xy,\text{ef}} = -\xi_{yx,\text{ef}}$. Furthermore, the global effective material response [Eq. (7a)] is described by tensors such that:

$$\overline{\varepsilon}_{\text{ef}} = \varepsilon_{\text{ef},\perp} \left(\hat{\mathbf{x}} \otimes \hat{\mathbf{x}} + \hat{\mathbf{y}} \otimes \hat{\mathbf{y}}\right) + \varepsilon_{\text{ef},zz} \hat{\mathbf{z}} \otimes \hat{\mathbf{z}}, \tag{15a}$$

$$\overline{\mu}_{\text{ef}} = \mu_{\text{ef},\perp} \left(\hat{\mathbf{x}} \otimes \hat{\mathbf{x}} + \hat{\mathbf{y}} \otimes \hat{\mathbf{y}}\right) + \mu_{\text{ef},zz} \hat{\mathbf{z}} \otimes \hat{\mathbf{z}}, \tag{15b}$$

$$\overline{\xi}_{\text{ef}} = \left(\overline{\zeta}_{\text{ef},\perp}\right)^T = \kappa_{\text{ef}} \left(\hat{\mathbf{x}} \otimes \hat{\mathbf{x}} + \hat{\mathbf{y}} \otimes \hat{\mathbf{y}}\right) - \xi_{\text{ef}} \hat{\mathbf{z}} \times \mathbf{1}. \tag{15c}$$

The above equations show that the homogenized material response is invariant under arbitrary rotations about the *z*-axis. Clearly, the effective material matrix $\mathbf{M}_{\text{ef}}$ is always symmetric and real-valued and thereby the response is Hermitian in the long wavelength limit. Evidently, the first piece of the magneto-electric tensors is a symmetric tensor that provides the axion (Tellegen) piece [50]. The axion response is isotropic in the *xoy* plane, and vanishes along the *z*-direction. On the other hand, the second term of $\overline{\xi}_{\text{ef}}$ provides an anti-symmetric moving-medium piece [50]. Both terms originate nonreciprocity in the long wavelength limit. A material with $\kappa_{\text{ef}} \neq 0$ and $\xi_{\text{ef}} \neq 0$ is known as a moving-Tellegen medium. In the next subsection, we will provide explicit formulas for all the effective parameters for a two-phase system and show how by tuning the angle $\theta$ it is possible to control the relative strength of $\kappa_{\text{ef}}$ and $\xi_{\text{ef}}$.

In summary, the glide rotation-symmetry [Eq. (12)] ensures that the effective response is independent of the field polarization in the *xoy* plane (see also the next subsection), and that the global response is coincident with that of a moving-Tellegen medium.

## C. Two-phase spacetime crystal

The glide-rotation symmetry constraint [Eq. (12)] can be implemented with a two-phase crystal such that the unit cell is formed by two layers *A* and *B* of identical thickness $d = a/2$ (see Fig. 1). The layer B is identical to layer A apart from a 90º rotation about the *z*-axis. This means that $\overline{\varepsilon}_{\perp,B} = \mathbf{J}^T \cdot \overline{\varepsilon}_{\perp,A} \cdot \mathbf{J}$ and $\overline{\mu}_{\perp,B} = \mathbf{J}^T \cdot \overline{\mu}_{\perp,A} \cdot \mathbf{J}$. These formulas can be written explicitly as a function of the permittivity and permeability components in the *xoy* coordinate system

$$\begin{pmatrix} \varepsilon_{11} & \varepsilon_{12} \\ \varepsilon_{12} & \varepsilon_{22} \end{pmatrix}_{\text{mat. A}} = \begin{pmatrix} \varepsilon_{22} & -\varepsilon_{12} \\ -\varepsilon_{12} & \varepsilon_{11} \end{pmatrix}_{\text{mat. B}}, \quad \begin{pmatrix} \mu_{11} & \mu_{12} \\ \mu_{12} & \mu_{22} \end{pmatrix}_{\text{mat. A}} = \begin{pmatrix} \mu_{22} & -\mu_{12} \\ -\mu_{12} & \mu_{11} \end{pmatrix}_{\text{mat. B}}, \quad (16)$$

where the subscripts "mat. A" and "mat. B" indicate if the permittivity and permeability components are calculated for material A or for material B, respectively. Clearly, the material parameters of layer A determine the entire geometry of the spacetime crystal.

Substituting the above material parameters in Eq. (10) one obtains after some manipulations the following analytical expressions for the effective parameters of the resulting a moving Tellegen bi-anisotropic material:

$$\varepsilon_{\text{ef},\perp} = \frac{\varepsilon_{11} + \varepsilon_{22} - (\varepsilon_{11}\varepsilon_{22} - \varepsilon_{12}^2)(\mu_{11} + \mu_{22})v^2}{2 - \frac{1}{2}(\varepsilon_{11} + \varepsilon_{22})(\mu_{11} + \mu_{22})v^2}, \quad \mu_{\text{ef},\perp} = \frac{\mu_{11} + \mu_{22} - (\mu_{11}\mu_{22} - \mu_{12}^2)(\varepsilon_{11} + \varepsilon_{22})v^2}{2 - \frac{1}{2}(\varepsilon_{11} + \varepsilon_{22})(\mu_{11} + \mu_{22})v^2} \quad (17a)$$

$$\kappa_{\text{ef}} = v\frac{(\varepsilon_{11} - \varepsilon_{22})\mu_{12} - (\mu_{11} - \mu_{22})\varepsilon_{12}}{2 - \frac{1}{2}(\varepsilon_{11} + \varepsilon_{22})(\mu_{11} + \mu_{22})v^2} \quad \text{and} \quad \xi_{\text{ef}} = -v\frac{\frac{1}{2}(\varepsilon_{11} - \varepsilon_{22})(\mu_{11} - \mu_{22}) + 2\varepsilon_{12}\mu_{12}}{2 - \frac{1}{2}(\varepsilon_{11} + \varepsilon_{22})(\mu_{11} + \mu_{22})v^2}. \quad (17b)$$

In the above formulas, all the permittivity and permeability elements are evaluated for layer $A$. Using Eq. (11) the effective parameters can also be written in terms of the eigenvalues of the permittivity and permeability tensors and of the $\theta$ angle as follows:

$$\varepsilon_{\text{ef},\perp} = \frac{\varepsilon_1 + \varepsilon_2 - \varepsilon_1\varepsilon_2(\mu_1 + \mu_2)v^2}{2 - \frac{1}{2}(\varepsilon_1 + \varepsilon_2)(\mu_1 + \mu_2)v^2}, \quad \mu_{\text{ef},\perp} = \frac{\mu_1 + \mu_2 - \mu_1\mu_2(\varepsilon_1 + \varepsilon_2)v^2}{2 - \frac{1}{2}(\varepsilon_1 + \varepsilon_2)(\mu_1 + \mu_2)v^2}. \quad (18a)$$

$$\kappa_{\text{ef}} = -\frac{v}{2}\frac{(\varepsilon_1 - \varepsilon_2)(\mu_1 - \mu_2)\sin 2\theta}{2 - \frac{1}{2}(\varepsilon_1 + \varepsilon_2)(\mu_1 + \mu_2)v^2} \quad \text{and} \quad \xi_{\text{ef}} = -\frac{v}{2}\frac{(\varepsilon_1 - \varepsilon_2)(\mu_1 - \mu_2)\cos 2\theta}{2 - \frac{1}{2}(\varepsilon_1 + \varepsilon_2)(\mu_1 + \mu_2)v^2}. \quad (18b)$$

Again, the eigenvalues of the permittivity and of the permeability are evaluated for the layer $A$. Interestingly, the transverse effective permittivity and permeability are independent of the angle $\theta$. Furthermore, $\kappa_{\text{ef}}^2 + \xi_{\text{ef}}^2$ is also independent of $\theta$. However, the relative strength of the axion and moving couplings is evidently controlled by the angle $2\theta$.

Figure 2b shows the effective parameters $\varepsilon_{\text{ef},\perp}, \mu_{\text{ef},\perp}, \kappa_{\text{ef}}, \xi_{\text{ef}}$ as a function of $\theta$, for a modulation speed $v = 0.2c$ in the subluminal regime. We choose a high contrast between the permittivity and permeability eigenvalues to make more evident the impact of the spacetime modulation. As seen, when the optical axes of the permittivity and permeability tensors are aligned, i.e., $\theta = 0, 90°, 180°$ the axion parameter vanishes, $\kappa_{ef} = 0$, resulting in an effective moving medium [41, 43, 44]. On the other hand, in agreement with Eq. (18b), it is possible to obtain a pure axion response by choosing $\theta = 45°, 135°$ so that $\xi_{ef} = 0$.

Figure 3 depicts the refractive index seen by the waves propagating along the +z-direction (solid line) and by the waves propagating along the –z-direction (dashed line) as a function of the angle $\theta$. Due to the invariance of the response under continuous rotations about the z-axis, the refractive index is independent of the wave polarization for all $\theta$. However, the refractive indices of counter-propagating waves are typically different, originating thereby a synthetic Fresnel drag in agreement with earlier works [41, 43]. Interestingly, for a pure axion ($\theta = 45°, 135°$) the refractive indices become identical and there is no Fresnel drag. The difference between the two refractive indices is maximized for $\theta = 0°, 90°$ which correspond to an ideal synthetic moving medium ($\kappa_{ef} = 0$).

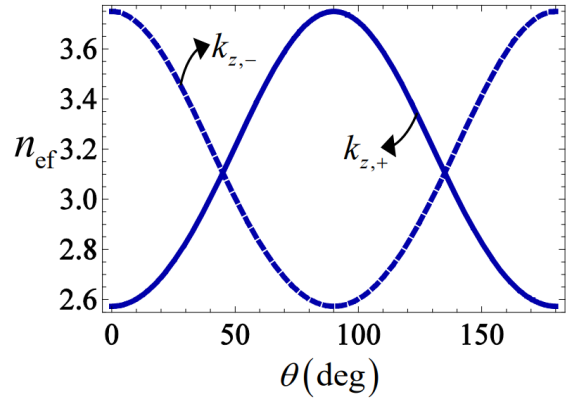

**Fig. 3** Refractive index $n_{ef}$ seen by the waves propagating along the +z-direction (solid line) and by the waves propagating along the –z-direction (dashed line) as a function of the angle $\theta$, for $\varepsilon_1 = 6$, $\varepsilon_2 = 2$, $\mu_1 = 3$ and $\mu_2 = 1$, and modulation speed $v = 0.2c$.

The refractive indices depicted in Fig. 3 are determined using the homogenization theory [Eq. (18)]. Interestingly, the homogenization result agrees exactly with the result obtained from the slopes of the dispersion of the Bloch modes of the spacetime crystal near the point $(\omega = 0, k_z = 0)$ (see Appendix C). This property will be further discussed in Sect. III.

In summary, the key ingredients required to have the (transverse-isotropic) axion response are the glide-rotation symmetry and an offset angle $\theta = 45°$ between the optical axes of the permittivity and permeability. In particular, a non-trivial axion response requires permittivity and permeability tensors with an anisotropic response in the *xoy* plane.

## III. THE SYNTHETIC AXION MEDIUM

Next, we investigate in detail the properties of the synthetic axion medium. Thus, in the rest of the article the offset angle is fixed at $\theta = 45°$. In such case, the effective medium [Eq. (15)] describes a uniaxial axion material such that $\overline{\xi}_{\text{ef}} = \overline{\zeta}_{\text{ef},\perp} = \kappa_{\text{ef}} \left( \hat{\mathbf{x}} \otimes \hat{\mathbf{x}} + \hat{\mathbf{y}} \otimes \hat{\mathbf{y}} \right)$.

It is convenient to write the eigenvalues of the permittivity and permeability tensors, $\varepsilon_1, \varepsilon_2$ and $\mu_1, \mu_2$, of layer *A* as follows:

$$\begin{aligned} \varepsilon_1 &= \varepsilon_{\text{M}}(1+\delta_\varepsilon), & \varepsilon_2 &= \varepsilon_{\text{M}}(1-\delta_\varepsilon) \\ \mu_1 &= \mu_{\text{M}}(1+\delta_\mu), & \mu_2 &= \mu_{\text{M}}(1-\delta_\mu) \end{aligned} \tag{19}$$

The parameters $\delta_\varepsilon, \delta_\mu$ determine the strength of the anisotropy of the electric and magnetic responses, respectively. The effective material parameters of the synthetic axion medium [Eq. (18) with $\theta = 45°$] can be written as:

$$\varepsilon_{\text{ef},\perp} = \varepsilon_{\text{M}} \frac{1 + \varepsilon_{\text{M}} \mu_{\text{M}} v^2 (\delta_\varepsilon^2 - 1)}{1 - \varepsilon_{\text{M}} \mu_{\text{M}} v^2}, \quad \mu_{\text{ef},\perp} = \mu_{\text{M}} \frac{1 + \varepsilon_{\text{M}} \mu_{\text{M}} v^2 (\delta_\mu^2 - 1)}{1 - \varepsilon_{\text{M}} \mu_{\text{M}} v^2}, \quad \kappa_{\text{ef}} = -\delta_\varepsilon \delta_\mu \frac{\varepsilon_{\text{M}} \mu_{\text{M}} v}{1 - \varepsilon_{\text{M}} \mu_{\text{M}} v^2}. \tag{20}$$

Equation (20) confirms that a nontrivial axion response requires that both the permittivity and permeability are anisotropic, i.e., $\delta_\varepsilon \neq 0$ and $\delta_\mu \neq 0$. If either $\delta_\varepsilon = 0$ or $\delta_\mu = 0$, the effective response becomes reciprocal. Note that when $\delta_\varepsilon = 0$ the permittivity of the two-phase crystal is uniform (i.e., there is no spacetime modulation of the permittivity) due to the assumed glide-rotation symmetry.

It is relevant to underline that even though the refractive index of the effective axion medium is polarization independent, each of the material phases is bi-refringent and reciprocal when on its own. Specifically, in the absence of time modulations, the velocity of the wave in layer A for propagation along the +z-direction depends on the polarization. The wave velocities for the two eigenpolarizations in material $A$ are (the same result is obtained for layer B):

$$\frac{v_\pm^{\text{aniso}}}{c} = \frac{1}{\sqrt{\varepsilon_M \mu_M} \sqrt{1 \pm \sqrt{\delta_u^2 + \delta_\varepsilon^2 - \delta_\varepsilon^2 \delta_u^2}}}. \qquad (21)$$

Thus, the sub-luminal range for the spacetime modulation is determined by:

$$0 < |v| < v_+^{\text{aniso}}, \qquad \text{(sub-luminal range)}. \qquad (22)$$

On the other hand, the super-luminal range is defined by the interval $|v| > v_-^{\text{aniso}}$. In this work, we focus on the sub-luminal regime wherein the spacetime crystal response is bounded and free of instabilities (there are no exponentially growing in time natural modes).

Equation (20) shows that the axion $\kappa_{\text{ef}}$ parameter is proportional to both $\delta_\varepsilon$ and $\delta_\mu$, which determine both the anisotropy and the modulation strength. Thus, $|\kappa_{\text{ef}}|$ increases always with $|\delta_\varepsilon|$ and $|\delta_\mu|$. This property is illustrated in Fig. 4a, for $\varepsilon_M = 4$ and $\mu_M = 2$ and for a fixed modulation speed $v/c = 0.2$. For the range of $\delta_\varepsilon$ and $\delta_\mu$ in the plot, the spacetime crystal is

–16–

always operated in the sub-luminal regime. For large $\delta_\varepsilon$ and $\delta_\mu$, the subluminal threshold ($v_+^{aniso}$) approaches the modulation speed ($v/c = 0.2$), and the axion response is greatly enhanced. For example, for $\delta_\varepsilon = \delta_\mu = 0.5$ the effective medium has a giant nonreciprocal axion response with $\kappa_{ef} \approx -0.59$. The corresponding effective (transverse) permittivity and permeability are $\varepsilon_{ef,\perp} \approx 4.47$ and $\mu_{ef,\perp} \approx 2.24$, respectively.

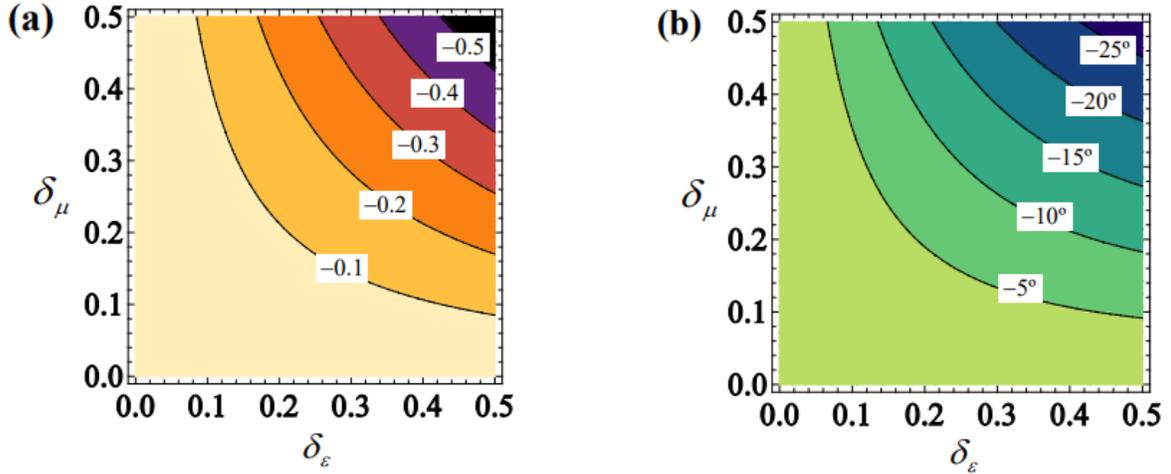

**Fig. 4 (a)** Density plot of the axion parameter $\kappa_{ef}$ as a function of $\delta_\varepsilon, \delta_\mu$. **(b)** Angle of rotation $\theta_R$ as a function of $\delta_\varepsilon, \delta_\mu$. In both plots, $\varepsilon_M = 4$ and $\mu_M = 2$ and $v/c = 0.2$.

Figure 5a shows the effective parameters of the synthetic axion medium as a function of the modulation speed $v$. As seen, for a non-trivial modulation speed the axion parameter becomes nonzero. Even for very moderate modulation speeds the axion parameter can be several orders of magnitude larger than what is observed in any natural material. In the subluminal range (pink shaded region), the sign of the axion parameter is locked to the sign of the modulation speed. The axion parameter flips sign in the transluminal region ($v_+^{aniso} < |v| < v_-^{aniso}$).

In agreement with Fig. 3, the effective refractive index seen by wave propagating in the synthetic axion medium is independent of the wave polarization and of the propagation (+z or –z) direction: $n_{ef} = \sqrt{\varepsilon_{ef,\perp}\mu_{ef,\perp} - \kappa_{ef}^2}$. One of the peculiar properties of the effective isotropic axion response is that the (instantaneous) electric and magnetic fields are not orthogonal, different from conventional dielectrics [8]. The offset angle $\phi$ with respect to 90º (see the inset of Fig. 5b) is given by:

$$\phi = \arcsin\left(\frac{\kappa_{ef}}{\sqrt{\varepsilon_{ef,\perp}\mu_{ef,\perp}}}\right). \tag{23}$$

Figure 5b depicts $\phi$ as a function of the modulation speed for the same structural parameters as before. As seen, for a velocity near the subluminal threshold ($v/c = 0.2$) the offset angle can be as large as $\phi \approx -11º$. The effective medium prediction of $\phi$ agrees very precisely with the calculation of $\phi$ based on the spatial averaging of the long-wavelength Bloch mode fields of the spacetime crystal. The Bloch modes are numerically determined as explained in Appendix C.

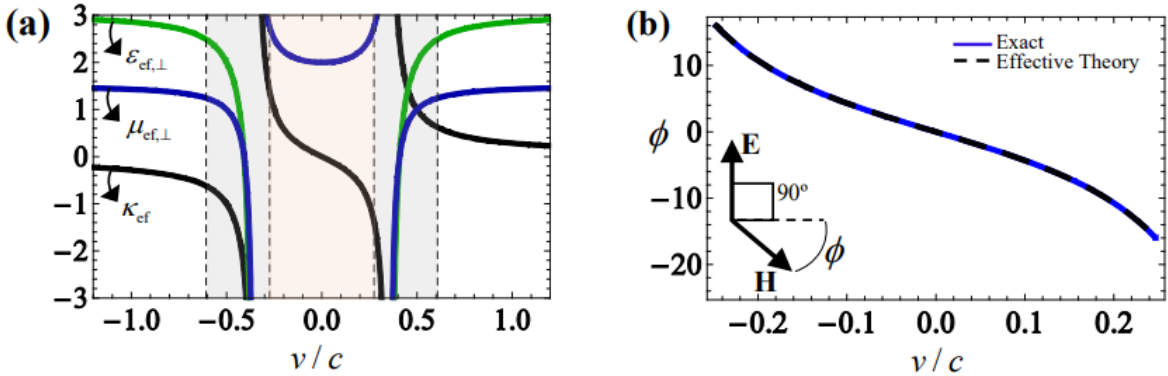

**Fig. 5 (a)** Effective parameters of the synthetic axion medium as a function of the normalized modulation speed. The dashed lines mark transitions between the sub-luminal range (pink shaded region), the transluminal range (grey shaded regions) and the super-luminal range (unshaded regions). **(b)** Offset angle $\phi$ between the magnetic and electric fields (see the inset) in the synthetic axion medium as a function of the modulation speed $v$. In both plots, $\varepsilon_M = 4$, $\mu_M = 2$ and $\delta_\varepsilon = \delta_\mu = 0.5$.

Furthermore, as shown in Fig. 6a, there is an excellent agreement between the effective refractive index $n_\text{ef}$ calculated with the homogenization theory (dashed line) and the refractive index extracted from the slope of the spacetime crystal dispersion in the lab frame in the long wavelength limit (solid line). This is further illustrated in Fig. 6b, which depicts the exact band diagram of the spacetime crystal for $v/c = 0.2$. The dashed lines represent the dispersion predicted by homogenization theory and match precisely the exact band structure for $\omega a/c \ll 1$, where $a$ is the lattice period. As previously mentioned, the slopes are symmetric in respect to $k_z$, i.e., $\omega(k_z) = \omega(-k_z)$, due to the isotropic response of the axion material. It should be noted that away from the long wavelength limit the condition $\omega(k_z) = \omega(-k_z)$ is violated and the nonreciprocity of the spacetime crystal is evident in the band diagram. For completeness, we also show the dispersion band diagram of the spacetime crystal in the Galilean co-moving frame (Fig. 6c). In the co-moving frame, the wave dispersion is asymmetric with respect to $k_z$ even in the long wavelength limit.

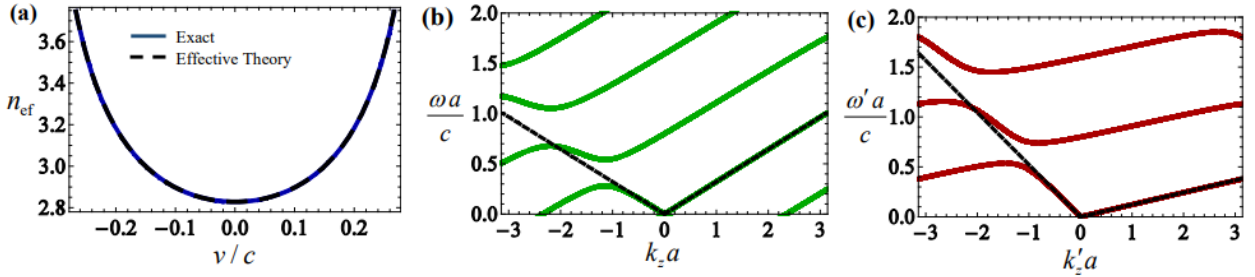

**Fig. 6 (a)** Equivalent refractive index $n_\text{ef}$ in the long wavelength limit as a function of the modulation speed $v$ calculated with (i) the slopes of the exact dispersion diagram of the spacetime crystal (solid line), and (ii) the effective medium theory (dashed line). **(b)** Band diagram in the lab frame of a spacetime crystal with the parameters $\varepsilon_\text{M} = 4$, $\mu_\text{M} = 2$, $\delta_\varepsilon = \delta_\mu = 0.5$ and $v/c = 0.2$. **(c)** Similar to (b) but calculated in the Galilean co-moving frame. The dashed black lines in (b)-(c) represent the effective medium dispersion.

In order to illustrate how the axion coupling can tailor the light-matter interactions in the spacetime crystal, we consider a simple scattering problem, where a normally incident linearly polarized plane wave propagating in air ($z<0$) illuminates the synthetic axion medium ($z>0$) (see the inset of Fig. 7). The wave scattering is characterized using the effective medium model.

As is well known [8], under the described conditions the reflected wave is also linearly polarized, but the electric field of the reflected wave is rotated by some angle $\varphi_R$ with respect to the incident field. The rotation of polarization of the reflected wave is a clear fingerprint of the axion response, and it is only possible due the nonreciprocity of the effective medium.

For an air-axion material interface the polarization rotation angle is given by [8]:

$$\varphi_R = \arctan\left(\frac{2\kappa_{\text{ef}}}{\varepsilon_{\text{ef},\perp} - \mu_{\text{ef},\perp}}\right). \tag{24}$$

It can be shown that $\varphi_R$ is independent of the axion-material slab thickness. Note that when $\varepsilon_{\text{ef},\perp} = \mu_{\text{ef},\perp}$ the angle is not defined because in that case the reflection coefficient vanishes (not shown). Figure 7 depicts the rotation angle $\varphi_R$ as a function of the modulation speed $v$. When $v/c = 0.2$ the angle can be as large as $\varphi_R \approx -27.76°$, resulting in a strongly nonreciprocal scattering rooted in the synthetic axion response.

The angle of rotation $\varphi_R$ may be written explicitly in terms of $\varepsilon_M, \mu_M, \delta_\varepsilon, \delta_\mu$ [Eq. (20)] as:

$$\varphi_R = \arctan\left(\frac{-2\delta_\varepsilon \delta_\mu \varepsilon_M \mu_M v}{\varepsilon_M - \mu_M + \varepsilon_M \mu_M v^2 \left[\mu_M\left(1-\delta_\mu^2\right) - \varepsilon_M\left(1-\delta_\varepsilon^2\right)\right]}\right) \tag{25}$$

A density plot of $\varphi_R$ as a function of $\delta_\varepsilon, \delta_\mu$ is shown in Fig. 4b. Similar to the axion parameter, the magnitude of $\varphi_R$ increases monotonically with $|\delta_\varepsilon|$ and $|\delta_\mu|$. For completeness, we point out that the combination of a chiral response with an axion-type response can be used to design one-way devices and optical isolators [12].

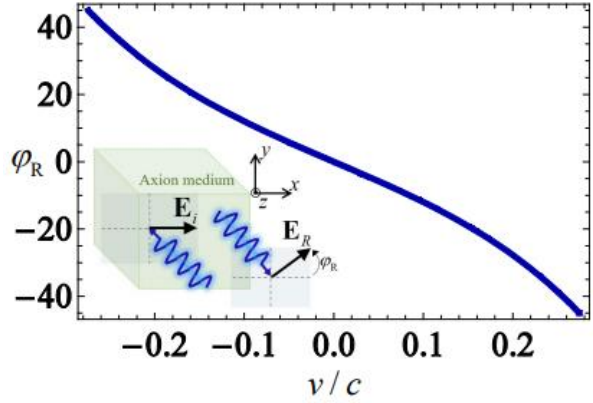

**Fig. 7** Angle of polarization rotation $\varphi_R$ for an air-synthetic axion medium interface, as a function of the modulation speed $v$. The parameters of the spacetime crystal are $\varepsilon_M = 4$, $\mu_M = 2$, and $\delta_\varepsilon = \delta_\mu = 0.5$. The angle $\varphi_R$ determines the rotation of the reflected field with respect to the incident linearly polarized electric field.

## IV. Conclusions

In summary, it was demonstrated that spacetime modulations provide an exciting path to sculpt the nonreciprocal response in the long wavelength limit. In particular, the type of anisotropy of the underlying photonic crystal (without time modulation) determines the symmetry of the effective magnetoelectric tensor of the metamaterial. While for isotropic materials the effective response reduces always to that of a moving medium, crystals made of anisotropic materials offer increased design flexibility and, in particular, may allow probing rather unique nonreciprocal couplings. Here, it was shown that when the spacetime crystal has glide-rotation symmetry its effective response in the *xoy* plane becomes independent of the wave polarization, and the effective material behaves as uniaxial moving-Tellegen medium. The relative strength of the axion and moving medium effective parameters depends on the angle $\theta$ between the optical axes of the permittivity and permeability tensors. It was shown that for a two-phase spacetime crystal an offset angle $\theta = 45º$ results in an ideal (uniaxial) synthetic axion response in the long wavelength limit. In particular, our work unveils a new physical mechanism and an exciting opportunity to implement the elusive Tellegen-medium response using spacetime modulations. It is envisioned that systems with a synthetic axion coupling may offer unique opportunities to sculpt and control nonreciprocal light-matter interactions.

**Acknowledgements**

This work was supported in part by the IET under the A F Harvey Engineering Research Prize, by the Simons Foundation Award Number 733700, and by Fundação para a Ciência e a Tecnologia and Instituto de Telecomunicações under Project No. UID/EEA/50008/2020.

## Appendix A: Material matrix transformation

A Lorentz transformation of coordinates preserves the structure of the Maxwell's equations provided the fields are transformed as (the primed frame moves with speed $\mathbf{v}$ with respect to the lab frame; $\gamma = 1/\sqrt{1-v^2/c^2}$ is the Lorentz factor) [51]

$$\mathbf{E}'_\| = \mathbf{E}_\|, \qquad \mathbf{B}'_\| = \mathbf{B}_\|, \tag{A1a}$$

$$\mathbf{E}'_\perp = \gamma\left(\mathbf{E}_\perp + \mathbf{v}\times\mathbf{B}\right), \qquad \mathbf{B}'_\perp = \gamma\left(\mathbf{B}_\perp - \frac{1}{c^2}\mathbf{v}\times\mathbf{E}\right), \tag{A1b}$$

and

$$\mathbf{D}'_\| = \mathbf{D}_\|, \qquad \mathbf{H}'_\| = \mathbf{H}_\|, \tag{A2a}$$

$$\mathbf{D}'_\perp = \gamma\left(\mathbf{D}_\perp + \frac{1}{c^2}\mathbf{v}\times\mathbf{H}\right), \qquad \mathbf{H}'_\perp = \gamma\left(\mathbf{H}_\perp - \mathbf{v}\times\mathbf{D}\right). \tag{A2b}$$

Here $\|$ and $\perp$ represent the field components parallel and perpendicular to the velocity. The same relations can be written in vector notation as:

$$\begin{pmatrix} \mathbf{E}' \\ \mathbf{H}' \end{pmatrix} = \underbrace{\begin{pmatrix} \gamma \mathbf{1}_\perp + \hat{\mathbf{u}}_\| \otimes \hat{\mathbf{u}}_\| & 0 \\ 0 & \gamma \mathbf{1}_\perp + \hat{\mathbf{u}}_\| \otimes \hat{\mathbf{u}}_\| \end{pmatrix}}_{\mathbf{A}} \begin{pmatrix} \mathbf{E} \\ \mathbf{H} \end{pmatrix} + \underbrace{\begin{pmatrix} 0 & \gamma \mathbf{v} \times \mathbf{1} \\ -\gamma \mathbf{v} \times \mathbf{1} & 0 \end{pmatrix}}_{\mathbf{V}} \begin{pmatrix} \mathbf{D} \\ \mathbf{B} \end{pmatrix} \quad \text{(A3a)}$$

$$\begin{pmatrix} \mathbf{D}' \\ \mathbf{B}' \end{pmatrix} = \frac{1}{c^2} \underbrace{\begin{pmatrix} 0 & \gamma \mathbf{v} \times \mathbf{1} \\ -\gamma \mathbf{v} \times \mathbf{1} & 0 \end{pmatrix}}_{\mathbf{V}} \begin{pmatrix} \mathbf{E} \\ \mathbf{H} \end{pmatrix} + \underbrace{\begin{pmatrix} \gamma \mathbf{1}_\perp + \hat{\mathbf{u}}_\| \otimes \hat{\mathbf{u}}_\| & 0 \\ 0 & \gamma \mathbf{1}_\perp + \hat{\mathbf{u}}_\| \otimes \hat{\mathbf{u}}_\| \end{pmatrix}}_{\mathbf{A}} \begin{pmatrix} \mathbf{D} \\ \mathbf{B} \end{pmatrix}, \quad \text{(A3b)}$$

with $\mathbf{1}_\perp = \mathbf{1} - \hat{\mathbf{u}}_\| \otimes \hat{\mathbf{u}}_\|$.

Suppose now that in the lab frame,

$$\begin{pmatrix} \mathbf{D}(x,y,z,t) \\ \mathbf{B}(x,y,z,t) \end{pmatrix} = \mathbf{M}(x,y,z,t) \cdot \begin{pmatrix} \mathbf{E}(x,y,z,t) \\ \mathbf{H}(x,y,z,t) \end{pmatrix}. \quad \text{(A4)}$$

Then, from Eq. (A3) it follows that in the co-moving frame:

$$\begin{pmatrix} \mathbf{E}' \\ \mathbf{H}' \end{pmatrix} = [\mathbf{A} + \mathbf{V} \cdot \mathbf{M}(x',y',z',t')] \cdot \begin{pmatrix} \mathbf{E} \\ \mathbf{H} \end{pmatrix}, \quad \begin{pmatrix} \mathbf{D}' \\ \mathbf{B}' \end{pmatrix} = \left[\frac{1}{c^2} \mathbf{V} + \mathbf{A} \cdot \mathbf{M}(x',y',z',t')\right] \cdot \begin{pmatrix} \mathbf{E} \\ \mathbf{H} \end{pmatrix}. \quad \text{(A5)}$$

Hence, the constitutive relations in the co-moving frame are:

$$\begin{pmatrix} \mathbf{D}' \\ \mathbf{B}' \end{pmatrix} = \left[\frac{1}{c^2} \mathbf{V} + \mathbf{A} \cdot \mathbf{M}\right] \cdot [\mathbf{A} + \mathbf{V} \cdot \mathbf{M}]^{-1} \cdot \begin{pmatrix} \mathbf{E}' \\ \mathbf{H}' \end{pmatrix}. \quad \text{(A6)}$$

so that the equivalent material matrix is given by:

$$\mathbf{M}' = \left[\frac{1}{c^2} \mathbf{V} + \mathbf{A} \cdot \mathbf{M}\right] \cdot [\mathbf{A} + \mathbf{V} \cdot \mathbf{M}]^{-1}. \quad \text{(A7)}$$

In the Lorentz transformation the parameter "$c$" can be regarded as a free-parameter not necessarily linked to the speed of light in vacuum $1/\sqrt{\varepsilon_0 \mu_0}$. The standard Lorentz transformation with $c = 1/\sqrt{\varepsilon_0 \mu_0}$ simply ensures that the vacuum response stays invariant under the coordinate transformation.

The case of Galilean transformation corresponds to taking the limit $c \to \infty$. In this limit, $\mathbf{A} = \mathbf{1}, \mathbf{V} = \begin{pmatrix} 0 & \mathbf{v} \times \mathbf{1} \\ -\mathbf{v} \times \mathbf{1} & 0 \end{pmatrix}$, and thereby:

$$\mathbf{M}' = \mathbf{M} \cdot [\mathbf{1} + \mathbf{V} \cdot \mathbf{M}]^{-1}. \tag{A8}$$

It can be shown (details omitted for conciseness) that the effective material response in the lab frame obtained using a Galilean transformation is exactly the same as the response obtained with the Lorentz transformation.

Let us now suppose that the permittivity and permeability tensors in the laboratory frame have the structure shown in Eq. (3). Then, $\mathbf{M}'$ determined by Eq. (A8) is of the form:

$$\mathbf{M}' = \begin{pmatrix} \varepsilon_0 \overline{\varepsilon}' & \dfrac{1}{c} \overline{\xi}' \\ \dfrac{1}{c} \overline{\zeta}' & \mu_0 \overline{\mu}' \end{pmatrix}, \qquad \text{with} \tag{A9a}$$

$$\overline{\varepsilon}' = \begin{pmatrix} \varepsilon'_{xx} & \varepsilon'_{xy} & 0 \\ \varepsilon'_{xy} & \varepsilon'_{yy} & 0 \\ 0 & 0 & \varepsilon_{zz} \end{pmatrix}, \qquad \overline{\mu}' = \begin{pmatrix} \mu'_{xx} & \mu'_{xy} & 0 \\ \mu'_{xy} & \mu'_{yy} & 0 \\ 0 & 0 & \mu_{zz} \end{pmatrix}, \tag{A9b}$$

$$\overline{\xi}' = \overline{\zeta}'^T = \begin{pmatrix} \xi'_{xx} & \xi'_{xy} & 0 \\ \xi'_{yx} & \xi'_{yy} & 0 \\ 0 & 0 & 0 \end{pmatrix}. \tag{A9c}$$

Note that $\mathbf{M}'$ is a symmetric real-valued matrix and thereby is diagonalizable. Clearly, two of the eigenvectors of $\mathbf{M}'$ are of the type $(\hat{\mathbf{z}} \quad 0)^T$ and $(0 \quad \hat{\mathbf{z}})^T$. The other 4 eigenvectors are of the type $(\mathbf{u}_E \quad \mathbf{u}_H)^T$ with $\mathbf{u}_E, \mathbf{u}_H$ some vectors in the *xoy* plane, i.e., perpendicular to the direction of the modulation speed. This means that the response of the crystal along the $z'$ direction (parallel to the modulation speed) is effectively decoupled from the response of the material in the *xoy* plane.

## Appendix B: Homogenization in the long wavelength limit

The effective medium theory links the spatially-averaged fields in the co-moving frame as:

$$\begin{pmatrix} \langle \mathbf{D}' \rangle \\ \langle \mathbf{B}' \rangle \end{pmatrix} = \mathbf{M}'_{ef} \cdot \begin{pmatrix} \langle \mathbf{E}' \rangle \\ \langle \mathbf{H}' \rangle \end{pmatrix}. \tag{B1}$$

Here, $\langle ... \rangle$ denotes the operation of spatial averaging. For a periodic structure the averaging is defined as:

$$\langle f \rangle = \frac{1}{a} \int_0^a dz' \, f(z'), \tag{B2a}$$

with *a* the lattice constant. As it is well known, due to the structure of the Maxwell's equations, for stratified systems such that $\mathbf{M}' = \mathbf{M}'(z')$ the tangential field components $\mathbf{E}, \mathbf{H}$, i.e., $E'_x, E'_y, H'_x, H'_y$, can be assumed constant in the long wavelength limit ($i\partial_{t'} = \omega' \to 0$). This means that

$$E'_x = \langle E'_x \rangle, \qquad E'_y = \langle E'_y \rangle, \qquad H'_x = \langle H'_x \rangle, \qquad H'_y = \langle H'_y \rangle. \tag{B3a}$$

Similarly, the normal components of the fields $\mathbf{D}, \mathbf{B}$ are also constant in the long wavelength limit:

$$D'_z = \langle D'_z \rangle, \qquad B'_z = \langle B'_z \rangle. \tag{B3b}$$

Taking into account that the microscopic fields are linked by $\begin{pmatrix} \mathbf{D}' \\ \mathbf{B}' \end{pmatrix} = \mathbf{M}'(z') \cdot \begin{pmatrix} \mathbf{E}' \\ \mathbf{H}' \end{pmatrix}$ with $\mathbf{M}'(z') = \mathbf{M} \cdot [\mathbf{1} + \mathbf{V} \cdot \mathbf{M}]^{-1}$ having the structure shown in Eq. (A9), it is clear from Eq. (B3) that the effective material response is of the type:

$$\mathbf{M}'_{\text{ef}} = \begin{pmatrix} \varepsilon_0 \overline{\varepsilon'_{\text{ef}}} & \dfrac{1}{c}\overline{\xi'_{\text{ef}}} \\ \dfrac{1}{c}\overline{\zeta'_{\text{ef}}} & \mu_0 \overline{\mu'_{\text{ef}}} \end{pmatrix}, \qquad \text{with} \tag{B4a}$$

$$\overline{\varepsilon'_{\text{ef}}} = \begin{pmatrix} \langle \varepsilon'_{xx} \rangle & \langle \varepsilon'_{xy} \rangle & 0 \\ \langle \varepsilon'_{xy} \rangle & \langle \varepsilon'_{yy} \rangle & 0 \\ 0 & 0 & \dfrac{1}{\langle 1/\varepsilon_{zz} \rangle} \end{pmatrix}, \qquad \overline{\mu'_{\text{ef}}} = \begin{pmatrix} \langle \mu'_{xx} \rangle & \langle \mu'_{xy} \rangle & 0 \\ \langle \mu'_{xy} \rangle & \langle \mu'_{xx} \rangle & 0 \\ 0 & 0 & \dfrac{1}{\langle 1/\mu_{zz} \rangle} \end{pmatrix}, \tag{B4b}$$

$$\overline{\xi'_{\text{ef}}} = \overline{\zeta'_{\text{ef}}}^T = \begin{pmatrix} \langle \xi'_{xx} \rangle & \langle \xi'_{xy} \rangle & 0 \\ \langle \xi'_{yx} \rangle & \langle \xi'_{yy} \rangle & 0 \\ 0 & 0 & 0 \end{pmatrix}. \tag{B4c}$$

Notice that the material parameters that determine the response in the *xoy* plane are simply averaged in space, whereas the material parameters in the direction of the modulation speed (*z*) are determined by the average of the inverse function.

The effective response in the lab frame can be found with an inverse Galilean transformation (compare with Eq. (A8)):

$$\mathbf{M}_{\text{ef}} = \mathbf{M}'_{\text{ef}} \cdot [\mathbf{1} - \mathbf{V} \cdot \mathbf{M}'_{\text{ef}}]^{-1}. \tag{B5}$$

It can be easily shown that the inverse Galilean transformation preserves the structure shown in Eq. (B4), so that one obtains Eq. (7) of the main text.

## Appendix C: Band structure of the spacetime crystal

In this Appendix, we explain how the exact band structure of the spacetime crystal can be calculated. For simplicity, we restrict our attention to waves propagating along the $z$-direction, i.e., along the direction of the modulation speed. In that case, the Maxwell's equations in the Galilean co-moving frame can be written as:

$$i\frac{d}{dz'}\boldsymbol{\sigma} \cdot \begin{pmatrix} \mathbf{E}'_\perp \\ \mathbf{H}'_\perp \end{pmatrix} = i\frac{\partial}{\partial t'}\left[ \mathbf{M}'_\perp(z') \cdot \begin{pmatrix} \mathbf{E}'_\perp \\ \mathbf{H}'_\perp \end{pmatrix} \right], \tag{C1a}$$

$$\mathbf{M}'_\perp(z') = \mathbf{M}_\perp(z') \cdot \left[\mathbf{1}_{4\times 4} + v\boldsymbol{\sigma} \cdot \mathbf{M}_\perp(z')\right]^{-1}. \tag{C1b}$$

In the above $\boldsymbol{\sigma}$ is defined as in Eq. (10), $\mathbf{M}_\perp(z')$ is given by Eq. (4), and $\mathbf{E}'_\perp = \begin{pmatrix} E'_x & E'_y \end{pmatrix}^T$ and $\mathbf{H}'_\perp = \begin{pmatrix} H'_x & H'_y \end{pmatrix}^T$ are the transverse fields in the co-moving frame. It should be noted that $\mathbf{M}'_\perp(z')$ is formed by the elements of $\mathbf{M}'(z')$ [Eq. (A8)] that determine the response of the crystal to a transverse excitation.

Let us find the Bloch mode solutions of Eq. (C1) with $i\frac{\partial}{\partial t'} = \omega'$ the oscillation frequency and $k'_z$ the Bloch wave number in the Galilean co-moving frame. We consider the case of a two-phase spacetime crystal [Fig. 1]. In each homogeneous region of space, Eq. (C1) is equivalent to $\frac{d}{dz'}\begin{pmatrix} \mathbf{E}'_\perp \\ \mathbf{H}'_\perp \end{pmatrix} = -i\omega'\boldsymbol{\sigma} \cdot \mathbf{M}'_\perp \cdot \begin{pmatrix} \mathbf{E}'_\perp \\ \mathbf{H}'_\perp \end{pmatrix}$. It was taken into account that $\boldsymbol{\sigma}^2 = \mathbf{1}_{4\times 4}$. This equation can be formally integrated as:

$$\begin{pmatrix} \mathbf{E}'_\perp(z') \\ \mathbf{H}'_\perp(z') \end{pmatrix} = \exp\left(-i\omega'(z'-z'_0)\boldsymbol{\sigma}\cdot\mathbf{M}'_\perp\right) \cdot \begin{pmatrix} \mathbf{E}'_\perp(z'_0) \\ \mathbf{H}'_\perp(z'_0) \end{pmatrix}, \tag{C2}$$

where $\exp(...)$ stands for the exponential of a matrix. Hence, for a two-phase crystal with layers *A* and *B* of identical thickness (half-lattice constant, $a/2$) one can write:

$$\begin{pmatrix} \mathbf{E}'_\perp(z'_0 + a) \\ \mathbf{H}'_\perp(z'_0 + a) \end{pmatrix} = \exp\left(-i\frac{\omega' a}{2}\boldsymbol{\sigma} \cdot \mathbf{M}'_{\perp,B}\right) \cdot \exp\left(-i\frac{\omega' a}{2}\boldsymbol{\sigma} \cdot \mathbf{M}'_{\perp,A}\right) \cdot \begin{pmatrix} \mathbf{E}'_\perp(z'_0) \\ \mathbf{H}'_\perp(z'_0) \end{pmatrix}. \quad (C3)$$

Here, $\mathbf{M}'_{\perp,i}$ is the (transverse) material matrix for layer *i=A, B*. Imposing the Bloch mode condition, $\begin{pmatrix} \mathbf{E}'_\perp(z'_0 + a) \\ \mathbf{H}'_\perp(z'_0 + a) \end{pmatrix} = e^{ik'_z a} \begin{pmatrix} \mathbf{E}'_\perp(z'_0) \\ \mathbf{H}'_\perp(z'_0) \end{pmatrix}$, one obtains the secular equation:

$$\det\left(\exp\left(-i\frac{\omega' a}{2}\boldsymbol{\sigma} \cdot \mathbf{M}'_{\perp,B}\right) \cdot \exp\left(-i\frac{\omega' a}{2}\boldsymbol{\sigma} \cdot \mathbf{M}'_{\perp,A}\right) - e^{ik'_z a}\mathbf{1}_{4\times 4}\right) = 0, \quad (C4)$$

whose solutions determine the dispersion $\omega'$ vs. $k'_z$ of the crystal in the Galilean co-moving frame. Note that for a fixed $\omega'$, the solutions for $k'_z$ are such that $e^{ik'_z a}$ is an eigenvalue of the matrix $\exp\left(-i\frac{\omega' a}{2}\boldsymbol{\sigma} \cdot \mathbf{M}'_{\perp,B}\right) \cdot \exp\left(-i\frac{\omega' a}{2}\boldsymbol{\sigma} \cdot \mathbf{M}'_{\perp,A}\right)$. The dispersion in the lab frame $\omega$ vs. $k_z$ can be found from the dispersion in the co-moving frame using $\omega = \omega' + k'_z v$ and $k_z = k'_z$.